\documentstyle[12pt]{article}
\def\degr{\hbox{$^\circ$}}
\def\arcmin{\hbox{$^\prime$}}
\def\arcsec{\hbox{$^{\prime\prime}$}}
\def\utw{\smash{\rlap{\lower5pt\hbox{$\sim$}}}}
\def\udtw{\smash{\rlap{\lower6pt\hbox{$\approx$}}}}

\def\fdg{\hbox{$.\!\!^\circ$}}

\def\farcs{\hbox{$.\!\!^{\prime\prime}$}}

\def\a{{$\alpha$}}
\def\l{{$\lambda$}}
\def\ctb{{CTB~80}}
\def\h{{$^{\rm h}$}}
\def\m{{$^{\rm m}$}}
\def\s{{$^{\rm s}$}}
\def\dd{{$\delta$}}
\def\ha{{\rm H$\alpha$}}
\def\hbeta{{\rm H$\beta$}}
\def\hnii{{\rm H$\alpha+[$NII$]$}}
\def\nii{{\rm $[$NII$]$}}
\def\sii{{\rm $[$SII$]$}}
\def\oii{{\rm $[$OII$]$}}
\def\oiii{{\rm $[$OIII$]$}}
\def\vel{{km s$^{-1}$}}
\def\et{{et al.}}
\def\flux{{$10^{-17}$ erg s$^{-1}$ cm$^{-2}$ arcsec$^{-2}$}}
\def\dens{{\rm cm$^{-3}$}}
\begin{document}

\title{The peculiar supernova remnant CTB 80}

\author{F. Mavromatakis$^1$,  J. Ventura$^{1,2,3}$,  E. V. Paleologou$^{2}$,
J. Papamastorakis$^{1,2}$}

\maketitle
University of Crete, Physics Department, P.O. Box 2208, 710 03 Heraklion, Crete, Greece

Foundation for Research and Technology-Hellas, P.O. Box 1527, 711 10 Heraklion, Crete, Greece

Max Planck Institut f\"ur Extraterrestrische Physik, D-85741 Garching, Germany

\begin{abstract}
Deep CCD exposures of the peculiar supernova remnant \ctb\ in the light of the 
\hnii, \sii, \oii, and \oiii\ filters have been obtained. 
These images reveal significant shock heated emission in the area of the 
remnant. 
An extended bright diffuse nebula in the south--east part of \ctb\ overlaps 
soft X-ray emission from ROSAT but it does not appear to be related to 
the remnant under study. 
New diffuse and filamentary structures are detected to the south,  
south -- east,  and north of PSR 1951$+$32 most 
likely associated with \ctb. Especially, the sulfur line image shows 
emission in the north along the outer boundary of the IRAS and HI shells. 
The [OIII] emission is filamentary suggesting shock velocities 
greater than 100 km s$^{-1}$ but its spatial extent is quite limited. 
Lower shock velocities are expected in the north and north--east areas 
of the remnant, since [OII] emission is present while [OIII] line 
emission is not detected.   
The comparison between the [OIII] and 
[OII] line images further suggest the presence of significant 
inhomogeneities in the interstellar medium. The flux calibrated images 
do not indicate the presence of incomplete recombination zones, and we 
estimate that the densities of the preshock clouds should not exceed 
a few atoms per cm$^{3}$. The typical projected angular widths of the 
observed filaments are $\sim$ 30\arcsec. 
Typical surface brightness values of the long [OIII] filament in the 
south are $\sim$ 12 $\times$ \flux\ while the [OII] image is characterized 
by fluxes of 10 -- 20 $\times$ \flux. 
The area covered by the optical radiation along with the radio emission 
at 1410 MHz suggest that \ctb\ occupies a larger angular extent than was
previously known.  
\end{abstract}
\vfill\eject
\section{Introduction}
Downes (\cite{dow70}), almost 30 years ago, proposed that the radio
source CTB~80 (G69.0+2.7)
was a supernova remnant. However, confirmation came only in 1974
when observations of total, and polarized intensity 
of a number of radio sources were performed by Velusamy and Kundu (\cite{vel74}).
The non-thermal emission, and the strong polarization of the source 
thus offered further support to the initial claim, and 
established the source's peculiar radio morphology.
The radio images show the presence of a ($\sim$ 1\arcmin) 
compact central source (spectral index
$\sim$ 0.0 for S$_\nu$ $\sim\ \nu^{-\alpha}$),
a plateau to the east of the compact core extending for
$\sim$ 10\arcmin $\times$ 6\arcmin\ (spectral index $\sim$ 0.3) and three 
large scale structures extending, roughly, to the north, to the east and to
the south--west (spectral index $\sim$ 0.8, Angerhofer \et\ \cite{ang81}, 
Strom and Stappers \cite{str00}). 
The source, depending upon the frequency of observation, displays
a different morphology while the polarization is quite strong over a
large part of the radio emission. 
Radio observations in 1984 and 1985 led Strom (\cite{
str87}) to propose that the radio properties of the central compact core 
indicated the presence of a central neutron star.
Interest in this source 
intensified following the discovery
of a 39.5 ms ($\sim 10^5$ yr old) radio pulsar at the center of the radio compact core
(Clifton \et\ \cite{cli87}). Shortly after the pulsar discovery, 
Fesen \et\ (\cite{fes88}) reported the detection of a shell of infrared 
emission correlated with the radio lobes of  \ctb. 
The IRAS shell centered $ 30 \arcmin $ east of the pulsar, and 
characterized by a diameter of 
$\sim$ 1\degr, is open in the south--west and has a higher surface brightness 
in the 60 $\mu$m wavelength regime than in the 100 $\mu$m regime. These properties 
led the authors to suggest that the presumably fast moving 
pulsar caught up the decelerating supernova shell and, through 
the pressure of its relativistic wind has broken off the
southwestern portion of the HI shell deforming the magnetic field structure, 
and producing the peculiar south--west protrusion in the radio emission.
This suggestion is further supported by the detection of an 
expanding HI shell by Koo \et\ (\cite{koo90}). The HI shell is clumpy 
(Koo \et\ \cite{koo93}), its south--west part is also open and is very well 
correlated with the infrared shell.   
The distance to \ctb\ is approximately 2 kpc, although distances 
in the range of 1.5 -- 2.5 kpc cannot be 
excluded (Koo \et\ \cite{koo93}, Strom \& Stappers \cite{str00}).
Major characteristics of the remnant in X-ray wavelengths  
are the low count rate, and its small spatial extent
(Wang and Seward \cite{wan84}, Safi--Harb \et\ \cite{saf95}).
The ROSAT data show that the soft X-ray flux is confined to the area
around the pulsar ($\sim$ 11\arcmin\ $\times$ 5\arcmin)
while the medium and hard X-ray flux show extended emission east, south--east 
of the pulsar. The spectrum indicates an average photon power-law
index of $\sim$ 1.8. The column density was not well determined due to the 
low X-ray counting statistics and was found to lie in the range of
10$^{21}$ -- 10$^{22}$ cm$^{-2}$
(Safi--Harb \et\ \cite{saf95}).
The authors also studied the temporal behavior of the central source and 
concluded that the pulsed fraction has a significant dependence on the energy.
\par
Several optical observations have been performed, though 
the majority of them have
focused on the area in the immediate neighbourhood of the pulsar
(e.g. Blair \et\ \cite{bla88}, Whitehead \et\ \cite{whi89}, and
Hester and Kulkarni \cite{hes89}). Wide field red plates 
taken by van den Bergh (\cite{ber80}) in \ha\ and \sii\ 
showed faint,
patchy nebulosity. Blair \et\ (\cite{bla84}) performed
interference filter observations in \hnii\ of two 
30\arcmin $\times$ 30\arcmin\ areas located
south--west and north--east of the pulsar. The area
 around the pulsar was
also observed in \oiii\ but no emission was detected apart from the
nebulosity surrounding the pulsar. Spectra taken at two restricted areas in
the south--west and the north--east indicate moderate absorption and
shock heating. 
\par
In this work we present the first comprehensive CCD imaging
in \hnii, \sii, \oii\ and \oiii, of the area around \ctb, covering a
wide field of $\sim$ 2\degr\ $\times$ 2\degr, at a resolution of
4\arcsec\ -- 5\arcsec. These new optical images reveal 
a rich network of filamentary and diffuse structures 
allowing for a better comparison with
existing radio, infrared, and X-ray data. 
Information about the observations and the data
reduction is given in Sect. 2, 
while in Sect. 3 we describe the
observed morphology in the various filters.  
Finally, in Sect. 4 we discuss the physical properties of the remnant, 
its positional relation, and interaction with the interstellar medium (ISM).
\section{Observations}
\subsection{Optical images}
The optical images of \ctb\ were obtained with the 0.3 m
telescope at Skinakas Observatory. The remnant was observed in July 9,
July 10, and October 13, 1999. Two different CCD's were used during
these observations. The first, was a 1024 $\times$ 1024 Thomson CCD
which resulted in a  69\arcmin\ $\times$ 69\arcmin\ field of view
and an image scale of  4\farcs12 per pixel.
The second was a 1024 $\times$ 1024 Site CCD which had a larger pixel size
resulting in a 89\arcmin\ $\times$ 89\arcmin\ field of view and an image
scale of 5\farcs21 per pixel. Our aim was to cover the whole field
defined by the faintest radio contours at 1410 MHz (Mantovani \et\
\cite{man85}). Given the large extent of the remnant, we performed several
pointings in order to have the best possible coverage of the field. At each
pointing the relevant field was observed for 1800 s at two different
instances, i.e. the total exposure time is 3600 s. Consequently,
overlapping areas have an effective exposure time of 7200 s.
Every field was projected to a common origin before any image
combination. The astrometric solutions were calculated with the aid of the
Hubble Space Telescope Guide Star catalogue.
The characteristics of the interference filters used during the
observations are listed in Table~\ref{filters}.
All coordinates in this work refer to epoch 2000.
\par
Standard IRAF and MIDAS routines were used for the reduction of the data.
Individual frames were bias subtracted and flat-field corrected using
well exposed twilight flat-fields. The spectrophotometric standard stars
HR7596, HR7950, HR8634,
and HR718 were used for flux calibration.
\section{Results}
\subsection{The \hnii\ and \sii\ line emission}
The field in the area of \ctb\ contains both filamentary and diffuse
structures. The nebulosity around the pulsar at \a\ $\simeq$ 19\h 53\s\ 
and \dd\ $\simeq$ 32\degr53\arcmin\ is clearly seen in Fig. ~\ref{fig1} 
as a slightly extended ``star-like'' object situated in a 
10\arcmin\ $\times$ 10\arcmin\ patch of diffuse \hnii\ emission. 
North of PSR 1951+32, several small scale  structures are detected in the
\sii\ image (Fig. ~\ref{fig2}). 
They are nicely correlated with the north radio ridge (Fig. ~\ref{fig3}) 
as well as
with the north part of the I(60 $\mu$m)/ I(100 $\mu$m) map of Fesen  \et\
(\cite{fes88}, hereafter FSS) while their
relation to the \ha\ emission (\sii\ / \ha\ $\sim$ 1.0) further 
suggests that we observe shocked material
associated with the remnant \ctb.
To the west of the pulsar and towards the south, we find the known
network of filaments (position I in Fig. ~\ref{fig1},
see also Fig. 1a of Blair \et\ \cite{bla84}).
In this area, diffuse emission is seen to the west of the bright filament
at \a\ $\simeq$ 19\h51\m\ and \dd\ $\simeq$ 32\degr32\arcmin\
(position II), lying just outside
the faintest radio contours at 1410 MHz (Mantovani \et\ \cite{man85}).
The typical surface brightness fluxes measured in the \hnii\ and \sii\ images 
lie in the range of 30--40 $\times$ \flux\ and 5--12 $\times$ \flux, 
respectively. 
South of the radio contours and at positions III and IV two patches
of emission are detected running along the west--east direction at
\dd\ $\simeq$ 32\degr10\arcmin. Faint emission, at the same declination,
appears to fill the gap between these two locations and might indicate their
physical relation. The optical radiation at location IV displays
filamentary structures of projected widths $\sim$ 10\arcsec\ to 30\arcsec.
The structure seems to curve towards the north, north--east. The
overall appearance of these structures in the \sii\ image is less
filamentary.
\par
The bright diffuse emission to the south--east of position IV occupies
an area of $\sim$ 30\arcmin\ $\times$ 18\arcmin\ and its relation to \ctb\
is not certain. A search in the SIMBAD database around
\a\ $\simeq$ 19\h55\m\ and \dd\ $\simeq$ 32\degr05\arcmin\ did not reveal
any previously known extended object, although diffuse emission is seen
in the POSS plates. North of this extended structure, we come across two
elongated structures at positions V and VI (Fig. ~\ref{fig1}, ~\ref{fig2}). 
The first one
is oriented in the SE to NW direction and is characterized by a length
of $\sim$ 20\arcmin\ and typical thickness of $\sim$ 30\arcsec\ while the
second is oriented along the SW to NE direction and extends for
$\sim$ 33\arcmin. The thickness of this structure is $\sim$ 1\arcmin\ and its
position coincides with the outer south part of the
I(60 $\mu$m)/ I(100 $\mu$m) map of FSS.
The positions of these locations (V and VI) 
completely overlap the radio
contours (eastern ridge) at 1410 MHz (Mantovani \et\ \cite{man85}).
At the NE boundary of the radio contours the known
Lynd's Bright Nebula 156 (Lynds \cite{lyn65}) is found, 
which is quite bright in both the \hnii\ and \sii\
filters. Part of LBN 158 is seen to the west of LBN 156 at the west edge
of our field.
Interestingly, at the north end of LBN 156 we see two thin arcs convex
to the west extending further to the north for $\sim$ 38\arcmin\
(location VII). These two arcs have a projected thickness of $\sim$ 30\arcsec\
and cross each other at \a\ $\simeq$ 19\h56\m30\s\ and
\dd\ $\simeq$ 33\degr50\arcmin, while the crossing 
area is rather
extended (10\arcmin\ $\times$ 4\arcmin).
The arcs are also seen in the \sii\ image but are less prominent.
\subsection{The \oiii\ \l 5007 \AA\ line emission}
The morphology of the area around \ctb\ in \oiii\ (see Fig. ~\ref{fig4})
is quite different
from that seen in the \hnii\ and \sii\ images.
The field is relatively clean of diffuse, patchy emission while
a few filamentary structures are seen.  The first filament
is detected SW of the pulsar and has a length of $\sim$ 3\arcmin\
and a projected thickness of $\sim$ 25\arcsec.
It is located at position I where the corresponding \hnii\ flux is lower
by a factor
of $\sim$ 2 than the flux of the neighbouring filaments.
Blair \et\ (\cite{bla84}) did not detect any \oiii\ emission in their
``SW'' spectrum since it was taken $\sim$ 11\arcmin\ away.
Another \oiii\ filament which has an \hnii\ and \sii\ counterpart is
located at \a\ $\simeq$ 19\h50\m30\s\ and \dd\ $\simeq$ 32\degr34\arcmin\
(our position II, position ``SSW'' of Blair \et\ \cite{bla84}).
Around \a\ $\simeq$ 19\h50\m\ and \dd\ $\simeq$ 32\degr15\arcmin\ we find a
filament which is $\sim$ 12\arcmin\ long and $\sim$ 40\arcsec\ wide (position
IIa in Fig. ~\ref{fig4}).
A noteworthy feature of this filament is its location along the outer
radio contours, in the south,
at 1410 MHz (Mantovani \et\ \cite{man85}). Emission
from the low ionization lines of \hnii, \sii\ and \oii\ 
(Fig. ~\ref{fig1},~\ref{fig2}, ~\ref{fig5}) is only
partially correlated with this \oiii\ emission.  
To the south of position IV a $\sim$ 17\arcmin\ long \oiii\ filamentary
structure is detected which has no obvious \hnii\ counterpart. Its thickness
ranges from $\sim$ 30\arcsec\ -- 60\arcsec.
This structure is $\sim$ 14\arcmin\ away to the
south of the outer 1410 MHz
radio contours.
A peculiar \oiii\ structure, which has a shape similar to the number ``9'',
is found around \a\ $\simeq$ 19\h53\m\ and \dd\ $\simeq$ 32\degr30\arcmin\
just inside the outer radio contour at 1410 MHz.
Finally, the last \oiii\ features present in the field are the arcs to the
north of LBN 156 (position VII).
However, only one of the arcs is seen for
$\sim$ 38\arcmin\ while the second one is seen for $\sim$ 10\arcmin.
\subsection{The low ionization line of \oii\ \l\l 3726, 3729 \AA}
The morphology of the field in \oii\ (Fig. ~\ref{fig5}) is generally 
similar to that seen
in \hnii\ and \sii\ but still, important diferences do exist.
Some of the filaments seen in \hnii, SW of the pulsar (position I), 
have a counterpart
in \oii\ while others do not. Furthermore, in the position of an \ha\
filament, in the same area,
we see in the \oii\ image localized emission areas scattered
along the general direction of the \hnii\ filament.
It is interesting to note that the filament seen in \oii\ in the south
of position I is better defined than in \hnii.
Additionally, we observe that the long \oiii\ filament (south of position IV) 
is seen as a series
of localized emission ``hills'' in \oii. The ``valleys'' (gaps)
between the bright ``hills'' have
typical lengths of the order of 0.2 -- 0.4 pc.
Also, while we detect \oii\ emission
at position IV there is practically no emission in \oiii.
The extended diffuse structure in the south is also visible in \oii\ 
while to its
north, we observe less structured emission which is partially correlated
with the \hnii\ emission (positions V and VI).
Finally, the bright nebula LBN 156 and the two arcs in the very north
of our field are present and well correlated with the \hnii\ and \sii\ line
emission. Deep exposures have been obtained from the area east of the two arcs 
and the results will be presented elsewhere (Mavromatakis \et\ \cite{mav00}). 
\par
An interesting aspect concerning almost all of our line images is that
the filaments in the west and the SW seem to define rather well the two
sides of a triangle much like the shape of any closed, outer contour at 1410 MHz
(Mantovani \et\ \cite{man85}). 
\subsection{The ASCA spectral data}
We have analyzed public ASCA data of \ctb\ in order to have an accurate
determination of the column density which would allow us to obtain estimates 
of the color excess, E(${\rm B-V}$) and the average ISM 
density. 
An estimate of the color excess is desirable since current measurements 
refer to the pulsar neighbourhood (e.g. Hester and Kulkarni \cite{hes89}) 
while away from the pulsar the measurements are less certain 
(Blair \et\ \cite{bla84}).  Determination of the hydrogen column density 
will provide us with an order of magnitude estimate of the 
ISM density around \ctb\ which can be compared with our
estimates of the preshock cloud densities.  
\par
The source was observed in June 17, 1993 (sequence id 50037010)
by the GIS and SIS detectors onboard the ASCA satellite. Applying the 
strict selection criteria  to the original data resulted in
25 ks and 19 ks exposure time for the GIS and SIS detectors, respectively. 
The hardness ratios  between the 0.8 - 2.0 keV, 2.0 - 6.0  keV and the
6.0 - 12.0 keV  bands did not reveal any statistically
significant spectral
variations. Consequently, all photons within a 8\arcmin\ radius from the
center of the GIS intensity distribution were accumulated to create a
source spectrum, while from the SIS data all photons within a rectangular 
area of $\sim$ 4\arcmin\ $\times$ 4\arcmin\ were accumulated. 
An important point to be mentioned concerns the background
subtraction. The available blank--sky background spectra are obtained from
observations towards high galactic latitudes. However, \ctb\ is only
$\sim$ 3\degr\ away from the galactic plane where a different
spectral behavior of the sky background is expected.
We have extracted a few observations
from the ASCA database which were performed in the direction of \ctb.
The analysis showed that a thermal (T $\sim$ 0.59 keV) and a non--thermal
component ($\alpha\ \sim$ 1.2) could fit the source free spectra close to
the galactic plane. Subsequently, these components were included in the fits
of the SNR spectra by keeping the temperature and power--law index fixed
at the forementioned values. 
\par
We find that a simple power-model attenuated by
the intervening column density provides a sufficient description of the data.
The photon power--law index is 1.95$^{+0.15}_{-0.11}$ and the column
density is 2.9$^{+0.5}_{-0.5}$ $\times$ 10$^{21}$ cm$^{-2}$. The quoted
errors refer to the 90\% probability level for two parameters of
interest ($\Delta \chi^2$ $=$ 4.61). In the case where we do not include
the galactic plane background component, we find
a photon power--law index of 1.88$^{+0.06}_{-0.05}$ and a column density
of 2.6$^{+0.3}_{-0.3}$ $\times$ 10$^{21}$ cm$^{-2}$. The two methods provide
results that agree within the statistical errors due to the low counting
statistics of \ctb. Nevertheless, the contribution of the galactic plane
background
should always be taken into account when analyzing source spectra near the
galactic plane. We note here that our results are in good agreement with the
results reported by Safi--Harb \et\ (\cite{saf95}), however, the allowed
range of the parameters is much narrower now.
\par
Current measurements of the color excess suggest values in the range of 
$\sim$ 0.7 -- 1 (e.g. Blair \et\ \cite{bla84}, \cite{bla88}). 
Assuming the relation
N$_{\rm H}$ = $6.8(\pm1.6)\times10^{21}$ $\times$ E(B-V) cm$^{-2}$ mag$^{-1}$
(Ryter \et\  \cite{ryt75}), and the best determined N$_{\rm H}$ from the
ASCA data, we derive a color excess of 0.38 $\pm$ 0.10 mag. Clearly, the
relation used is a statistical relation and deviations of 3 -- 4$\sigma$
could be justified. However, if the observed difference is true, then the 
X-ray measured color excess of $\sim$ 0.4  will represent the amount of pure 
interstellar extinction. The additional 0.3 -- 0.4 mags of extinction would be 
due to local extinction around the core area. 
In any case, applying the interstellar reddening curve of Whitford
(\cite{whi58}) as presented by Kaler (\cite{kal76}), an
interstellar extinction c of 0.57 ($\pm$ 0.15) is obtained.
Finally, the average ISM density towards \ctb\ is 
$\sim$ 0.5 \dens, assuming a distance of 2 kpc 
(Koo \et\ \cite{koo93}, Strom and Stappers \cite{str00}). 
\section{Discussion}
\subsection{The HI and infrared shells}
FSS reported the detection of an infrared emitting shell 
in the area of \ctb. The shell posseses an angular diameter of $\sim$ 1\degr, 
is characterized by strong 60 $\mu$m emission relative to the 100 $\mu$m 
emission and is correlated with the radio emission. 
HI observations by Koo \et\ (\cite{koo90}) revealed the presence of a clumpy 
shell, open in the south--west, expanding at $\sim$ 70 km s$^{-1}$. 
They also estimated a density of the HI medium of 1.7 cm$^{-3}$. 
Subsequent VLA HI observations by Koo \et\ (\cite{koo93}) showed that the 
clumpy nature of the HI shell is caused by dense, fast moving HI clumps. These 
clumps preexisted the supernova explosion and were accelerated by the blast wave.  
The two shells (infrared and HI) are very well correlated suggesting their 
association with \ctb, while the smaller extent of the HI shell is probably due 
to the difficulty in distinguishing emission of the low radial 
velocity portion of the HI shell from galactic background 
emission (Koo \et\ \cite{koo93}).
\par
The optical emission detected north, north--east of the pulsar position 
is rather well correlated with the IRAS
\footnote{The digital infrared data, originally presented by Fesen 
\et\ (\cite{fes88}), were retrieved from http://skyview.gsfc.nasa.gov} 
ratio map of I(60 $\mu$m)/I(100 $\mu$m) of FSS as well
as with the HI shell of Koo 
\et\ ({\cite{koo90}).
The optical emission in this area lies inside the outer infrared (dashed) 
contour and outside the HI (dash--dotted) contour in Fig. \ref{fig6}. 
The solid contour corresponds to radio continuum emission at 49 cm 
(Strom \& Stappers \cite{str00}) while the three short lines in the south, 
south--west indicate the positions of the prominent \oiii\ filaments.  
As seen from Fig. \ref{fig6} there is also a good degree of 
correlation between HI, infrared and optical emission in the south, 
south--east (positions V and VI). 
The degree of correlation between the optical and infrared data increases 
when we consider the individual IRAS maps of e.g. the 60 $\mu$m emission. 
Optical forbidden line emission at positions IV and V  
seem to be related to the 60 $\mu$m 
emission. Based on this positional correspondence between the optical, 
HI and infrared emission, and the estimated flux ratio of 
[SII]/\ha\ $\sim$ 0.7 -- 1.0 (extinction of $\sim$ 0.6 asssumed) 
at these areas, it is proposed that the 
observed optical emission is indeed 
associated with \ctb.
The filamentary and diffuse structures in locations III, IV, V, and VI
along with LBN 156 may be viewed as forming part of an,
almost continuous, shell structure. This
adds further support to FSS's previous suggestion that LBN 156 is 
also associated with \ctb.
\subsection{The south--west area}
The calibrated flux images at positions I and II (Fig. ~\ref{fig1})
do not provide any strong evidence for incomplete
recombination zones. The estimated values of \oiii\ / \hbeta\
are found in the range of 1 -- 3 while a value of $\sim$ 7 (the minimum 
extinction of 0.6 was assumed) is obtained
at position IIa (Fig. ~\ref{fig4}). Raymond \et\ (\cite{ray88})
report a maximum ratio of 6 for complete shock structures and shock
velocities between 80 and 140 km s$^{-1}$. Since we are dealing most likely
with complete shocks, the results of
Cox and Raymond (\cite{cox85}) and Raymond \et\ (\cite{ray88}) can be used
to infer some basic shock properties. The presence of faint
\oiii\ emission ($\sim$ 6 $\times$ \flux) together with the high values
of the [OII]/[OIII] ratio ( $>$ 3--8, extinction of $\sim$0.6 asssumed) 
suggest shock velocities between
85 and 120 km s$^{-1}$, and most probably towards the high end of this range.
Since the column density of a fully developed
flow, roughly, scales as 10$^{18.5}\ \times$ u$_{s,100}^2$ cm$^{-2}$
(Raymond \et\ \cite{ray88}), where
u$_{s,100}$ denotes the shock velocity in units of 100 km s$^{-1}$,
we can set upper limits to the preshock cloud density.
The column density is
${\rm N =}$ $ \int_0^l n(l)\, dl \ > $ ${\rm n_c l,}$ where l is the
projected thickness of a filament and ${\rm n_c}$ the preshock cloud density.
A typical thickness of a filament at positions I and II is 9 $\times$
10$^{17}$ cm for a distance of 2 kpc 
(Koo \et\ \cite{koo93}, Strom and Stappers \cite{str00}),
thus we obtain ${\rm n_c}$ $<$ 3.6 $\times$ u$_{s,100}^2$ \dens.
For shock velocities in the range of 85 and 120 km s$^{-1}$, we get
${\rm n_c}$ $<$ 2.6 \dens\ and 5.2 \dens, respectively.
\par
In neighbouring locations around positions I and II possible \oiii\ emission is
below our detection limit while \oii\ emission is detected well above the
sky background. Since these two lines originate from the same element, it
is rather likely that the shock velocity is less than 100 km s$^{-1}$.
In this case, the upper limit to the preshock cloud density
becomes 3.6 \dens. A possible explanation for the lower shock velocities
could involve higher cloud densities by factors of 1.5 -- 2.0 than the
cloud densities where both \oii\ and \oiii\ are detected.
A different situation seems to hold for the \oiii\ filament at position IIa.
It is partially correlated with \oii\ emission while the degree of correlation
is even smaller when the \hnii\ and \sii\ images are examined.
These morphological differences suggest the presence of
inhomogeneities in the preshock medium. The low ionization lines are
produced in areas of lower temperatures while the \oiii\ line
emerges from higher temperature regions.
The less hot areas are found at larger distances behind the shock front
and consequently, at higher column densities.
Inhomogeneities in the preshock clouds would mainly affect regions of
higher column densities. At the same time, preshock cloud density
variations would also affect the recombination zone through their effect
on the cooling and recombination time scales (see Hester \cite{hes87} for
more details).
\subsection{The bright \oiii\ filament in the south}
In the south and close to position IV, a long \oiii\ filament
is detected. Typical surface brightness values are of the order of
12 $\times$ \flux. The same spatial location in the \oii\ image
looks completely different where small size patches ($\sim$ 50\arcsec\ $\times$
30\arcsec) of emitting material are seen, instead of a well defined filament.
According to Cox and Raymond (\cite{cox85}) shock velocities greater
than 100 km s$^{-1}$ would be required to give rise to the
observed \oii\ and \oiii\ fluxes.
Inhomogeneities and preshock density variations are probably the cause of the
observed differences in morphology. The CCD images suggest angular
length scales of $\sim$ 25\arcsec\ of these irregularities.
An important issue 
related to the presence of this filament is whether or not it is 
associated with \ctb. Even though, a negative correlation can not be formally 
excluded, a positive correlation may be more favorable given the shape, 
the spatial location and emission characteristics of this \oiii\ filament. 
In addition, we note that this filament is located along the outer boundary 
of a region of relatively strong 60 $\mu$m emission seen in the IRAS maps 
(Wheelock \et\ \cite{whe94}).  
A positive indentification would imply a larger extent than
currently assumed and, consequently, a larger shock radius.
New, more sensitive radio observations at 1410 MHz would help to 
clarify this issue.
\subsection{The east area of CTB 80}
The east, south--east area of \ctb\ shows mainly diffuse optical radiation,
eventhough a few filamentary structures are present. These are characterized 
by lengths of $\sim$ 60\arcsec\ -- 110\arcsec\ and projected widths of 
$\sim$ 20\arcsec. Assuming complete shock structures, shock velocities less 
than 100 km s$^{-1}$ and a distance of 2 kpc (Koo \et\ \cite{koo93}, 
Strom and Stappers \cite{str00}), 
we find that the preshock 
cloud densities n$_{\rm c}$ should satisfy n$_{\rm c}$ $<$ 5 cm$^{-3}$. 
Such shock velocities may be expected, according to the calculations 
of Cox and Raymond (\cite{cox85}), since associated [OIII] emission is not 
detected, while in one location (\a\ = 19\h55\m35\s\ and 
\dd\ = 32\degr33\arcmin03\arcsec) [OII] is also detected well above the 
sky background.  
Safi--Harb \et\ (\cite{saf95}) analyzed ROSAT PSPC and HRI data taken from  
the field of \ctb. The PSPC data show the presence of four extended irregular 
structures in the south--east apart from the emission observed around the 
pulsar. The whole enhancement is described as the cone--like feature 
by Safi--Harb \et\ \cite{saf95}) and includes an elongated structure  
along the outer boundary of the PSPC in the south--east.
The cone--like feature 
overlaps the optical emission at positions V and VI and the elongated 
structure partially overlaps the bright diffuse emission south of position IV. 
Given the strong sulfur line emission observed from this bright diffuse 
structure, it would be interesting to see if future radio observations 
would detect a thermal or non--thermal spectrum.  
However, at the moment, no firm conclusions can be drawn about the 
correlation between the optical and X-ray data.
\subsection{The nature and extent of \ctb}
Though shell-like structures have been seen in HI, the infrared, and now
in the optical, \ctb\ is no common shell SNR. There is substantial
and multifaceted evidence that PSR 1951+32 has strongly interacted
with and altered the remnant as is evident from its unusual radio
appearance. It is necessary to proceed with care in attempting to
reconstruct its evolution. As noted earlier (Sect. 4.2) the optical
observations appear to support the suggestion of FSS and Koo \et\
(\cite{koo90}) for a breakout of the SN shell south west of the pulsar,
presumably through the action of a strong particle and electromagnetic
wind, which has deformed the shell's magnetic field, and has given rise
to the south western radio emission ridge.
\par
The radio images at 1410 MHz and 1720 MHz show that faint emission extends 
for more than 1\fdg5 in the north--south and west--east directions. 
The optical data span in declination 
from 32\fdg2 to 33\fdg5, while the span in right ascension is 
somewhat smaller. 
The new optical emission features detected in the south, south--east 
(positions IV, V and VI) seem to define an extended emission arc of large 
curvature, while the optical features detected in the north delinetate an 
emission arc of smaller curvature. 
This apparent discrepancy could be reconciled if we adopt 
the evolutionary scenario of \ctb\ proposed by Koo \et\ (\cite{koo93}). 
According to this scenario, soon after the supernova explosion ($\sim$ 2000 yr) 
the blast wave broke out into a cavity in the south--east, while in the north, 
north--west the shock front propagated into a denser but clumpy interstellar 
medium. 
The pulsar was born with a westward velocity, and caught up the slower 
expanding shell after $\sim$ 10$^5$ yr. Consequently, the optical structures 
at positions III, IV and VI may roughly define the south, south--east boundaries 
of a cavity seen in the HI data, while the structure at position V may be 
projected into the cavity area, or else be related to the cavity's 
foreground edge. 
\par
If this interpretation is correct, then the 
optical data suggest that the cavity must extend from declination 
$\simeq$ 33\degr10\arcmin\ to $\simeq$ 32\degr10\arcmin.  
The correlation between the enhanced IRAS emission along the east cavity 
wall, and the optical radiation may suggest that the shock front encountered 
a dense medium. Note here that diffuse X--ray emission is mainly detected 
in this area (Safi--Harb \et\ \cite{saf95}) which may tie with the 
presence of the HI cavity permeated by hot, low density gas 
(Koo \et\ (\cite{koo93}).
The secondary shock driven into these dense clouds 
allowed the production of \ha, \nii, and \sii\ radiation but its velocity 
was not high enough to allow for the production of \oiii\ radiation, i.e. 
the shock velocity is probably less than $\sim$ 70 \vel\ 
 (Cox \& Raymond \cite{cox85}, 
Hartigan \et\ \cite{har87}).  However, \oiii\ filamentary emission 
(positions I, II, IIa, and III) is clearly detected in the south, 
south-west (Fig. \ref{fig4}, Fig. \ref{fig6}). Consequently,  
shock velocities greater than $\sim$ 80 \vel\ are expected in these locations
according to the calculations of Cox \& Raymond (\cite{cox85}). 
These areas are found to the south of the dense HI cloud detected by 
Koo \et\ (\cite{koo93}) where the IRAS and HI emission is weak (FSS, 
Koo \et\ \cite{koo90}). 
The presence of shock heated filaments at positions I and II suggests 
that the injection of pulsar--generated relativistic particles 
(FSS) has a strong impact on the ``break-out'' portion of the remnant's 
south western segment which may have been compressed westward, 
possibly against preexisting ``clouds''. 
The rich network of filaments seen in the low ionization lines and 
the spatially limited but filamentary \oiii\ emission (positions I and II) on 
the leading edge of the south--west radio ridge indicate different 
shock velocities as well as inhomegenities in the interstellar medium (see 
also Koo \et\ \cite{koo93}). 
\par
Summarizing we can propose the following. 
The optically emitting gas detected in the north in the \sii\ filter 
delineates a circular structure which is very well correlated with the 
radio, infrared and HI radiation. 
The north part of the emission corresponds to the 
undisturbed part of the shock while in the south--east the 
accelerated shock front has reached the oposite walls of the HI cavity.
The shock velocity into the cavity walls was sufficiently low not to 
allow for the production of \oiii\ emission while faster shocks have 
propagated towards the south, south--west. The pulsar relativistic wind 
interacts with the preexisting clouds giving rise to a complex network
of filaments along the south--west radio ridge. 
\section{Conclusions}
A $\sim$ 2\degr$\times$2\degr\ area around the pulsar PSR 1951$+$32 was 
observed in \hnii, \sii, \oiii\ and \oii.
The low ionization line images of \sii\ and  \oii\ show significant emission 
not detected before at locations in the south and south--east. 
The \sii\ image shows also emission north of the pulsar, clearly, tracing 
the infrared shell and the north radio ridge. 
The morphology of the \hnii\ and \sii\ images is similar. 
The \oiii\ image appears rather faint in relation to the images in the lower
ionization lines, probably reflecting the remnant's advanced age and low
expansion velocity. 
The differences between the oxygen line images and the morphologies seen 
in the lower ionization images suggest the existence of density variations 
and inhomogenities in the preshock interstellar medium. Rough upper limits 
to the preshock density of the interstellar clouds are of the order of a 
few atoms per cm$^3$ while the ASCA X-ray data suggest an average insterstellar 
medium density of $\sim$ 0.5 \dens. Based on our upper limits, it is estimated that the 
density contrast between the ISM clouds and the ISM medium is less than 
$\sim$ 5. 
\par
The absence of radio emission from the main body of the infrared
and HI expanding shells is conspicuous and surprising, especially
as these shells are reported to be very massive 
(Koo \et\ \cite{koo90}, \cite{koo93}).
The area covered by the 1410 MHz radio emission and the optical 
radiation detected in the south and south--east indicate 
that the accelerated shock front has traversed the HI cavity 
and is interacting with its walls.  
More sensitive radio observations at 1410 MHz would be required to 
establish the relation of the optical structures in the south of 
\ctb, and the actual size of the remnant which is a crucial parameter 
for the further modeling of this remnant.  
An extended diffuse structure is detected in the south--east 
well away from the faintest radio contours which overlaps soft X-ray emission 
detected by ROSAT. Radio observations would be required to establish the 
nature of this structure.  
}

\begin{table}
      \caption[]{Interference filter characteristics}
         \label{filters}
\begin{flushleft}
\begin{tabular}{lllll}
            \hline
            \noalign{\smallskip}
Filter     & Wavelength$^{\rm a}$	& Line (\%)	\cr
           & (FWHM)(\AA)& contributions       \cr
            \hline
H\a\ +[NII] & 6555 (75)	& 100, 100, 100$^{\rm b}$	\cr
            \hline
[SII]       & 6708 (27)	& 100, 18$^{\rm c}$	\cr
            \hline
[OII]      & 3727 (28)	& 100, 100$^{\rm d}$		\cr
	    \hline
[OIII]      & 5005 (28)	& 100		\cr
            \hline
Cont red    & 6096 (134)&  --      \cr
		\hline
Cont blue   & 5470 (230)  & --     \cr
		\hline
\end{tabular}
\end{flushleft}
${\rm ^a}$ Wavelength at peak transmission for $f$/3.2\\\
${\rm ^b}$ Contributions from $\lambda$6548, 6563, 6584 \AA \\\
${\rm ^c}$ Contributions from $\lambda$6716, 6731 \AA \\\
${\rm ^d}$ Contributions from $\lambda$3727, 3729 \AA
   \end{table}
\newpage
  \begin {figure*}
    \caption{CTB~80 imaged in the \hnii\ filter.
     The image has been smoothed to suppress the residuals from the imperfect
     continuum subtraction. Here and in all subsequent figures north is up, 
     east to the left and the coordinates refer to epoch 2000.
     Shadings run linearly from 0.0 to 70 $\times$ \flux. 
     The line segments seen near overexposed stars in this figure and the 
     next figures are due to the blooming effect. }
     \label{fig1}
  \end{figure*}
  \begin {figure*}
    \caption{CTB~80 imaged in the [SII] filter.
     The image has been smoothed to suppress the residuals from the imperfect
     continuum subtraction.
    Shadings run linearly from 0.0 to 20 $\times$ \flux.}
     \label{fig2}
  \end{figure*}
\begin {figure*}   
    \caption{The 49 cm radio contours (Strom and Stappers \cite{str00}) 
 are overlaid to the [SII] image 
 shown in Fig. 2. The correlation of the small scale structures
 in the north is evident. The known filaments in the south--west are also 
 seen. The radio contours are drawn linearly 
 from 0.004 to 0.14 Jy$/$beam.}
     \label{fig3}
  \end{figure*}
  \begin {figure*}
    \caption{CTB~80 imaged in the \oiii\ \l 5007 \AA\ emission line.
     The image has been smoothed to suppress the residuals from the imperfect
     continuum subtraction. Shadings run linearly from 
     0.0 to 10 $\times$ \flux.}
     \label{fig4}
  \end{figure*}
  \begin {figure*}
    \caption{CTB~80 imaged in the \oii\ \l 3726 and 3729 \AA\ emission lines.
     The image has been smoothed to suppress the residuals from the imperfect
     continuum subtraction. Shadings run linearly from 
     0.0 to 24 $\times$ \flux.}
     \label{fig5}
  \end{figure*}
  \begin {figure*}
    \caption{CTB~80 imaged in the \sii\ filter as a grey scale representation. 
    The solid line outlines the 49 cm radio emission and the arrow points to 
    the pulsar location. The dashed line delineates the infrared shell while the
    dash--dotted line delineates the HI shell. The short lines in the south 
    labeled as [OIII]a, b, c indicate the locations of \oiii\ filaments.  
     }
     \label{fig6}
  \end{figure*}

\end{document}